\documentclass[aps,prb,twocolumn,showpacs, floats]{revtex4}

\usepackage{amsmath,amssymb,dsfont,graphicx,psfrag}
\newcommand{\nn}{\nonumber}

\begin{document}

\title{Quantum criticality in itinerant metamagnets}
\author{Mario Zacharias}
\author{Markus Garst}
\affiliation{Institut f\"ur Theoretische Physik, Universit\"at zu K\"oln, 50938 K\"oln, Germany}

\date{\today}
\begin{abstract} 
Critical thermodynamics close to a metamagnetic quantum critical endpoint (QCEP) in a metal is discussed within the framework of spin-fluctuation theory. We analyze the effective potential for the Ising order parameter 
that is renormalized by spin-fluctuations and acquires a characteristic temperature dependence. The resulting magnetic field, $H$, and temperature, $T$, dependence of the magnetization, specific heat, thermal expansion, magnetostriction, susceptibility and the Gr\"uneisen parameter are determined, and the crossover lines in the $(H,T)$ phase diagram are specified. We point out that the metamagnetic QCEP is intrinsically unstable with respect to a magnetoelastic coupling.
\end{abstract}
\pacs{ }
\maketitle
\section{Introduction}
In metamagnets the magnetization exhibits a superlinear rise as function of the applied magnetic field $H$ or even jumps 
at a first-order transition.\cite{Wohlfarth62} In the $(H,T)$ plane such a line of first-order metamagnetic transitions terminates in a critical endpoint $(H^*,T^*)$ beyond which metamagnetism reduces to a crossover phenomena. Interestingly, an isolated quantum critical endpoint (QCEP) arises if the endpoint temperature $T^*$ can be tuned to zero by some external control parameter. 

The notion of a metamagnetic QCEP was introduced in the context of the layered perovskite Sr$_3$Ru$_2$O$_7$ (Ref.~\onlinecite{Perry01, Grigera01}) to account for its peculiar metamagnetic properties. Here, the line of first-order metamagnetic transitions can be tuned to zero either using the orientation of the magnetic field \cite{Grigera03} or by applying pressure.\cite{Wu11} In the former case, however, the putative QCEP is preempted by the formation of a nematic phase,\cite{Grigera04,Stingl11} see Ref.~\onlinecite{Mackenzie12} for a review. 
Subsequently, the metamagnetic crossover in the single-layer ruthenate Ca$_{2-x}$Sr$_x$RuO$_4$ (Ref.~\onlinecite{Baier06}) as well as the heavy-fermion system CeRu$_2$Si$_2$ (Refs.~\onlinecite{Paulsen90, Weickert10}) was discussed in terms of metamagnetic quantum criticality motivated by the thermal expansion that is similar for the three compounds.\cite{Gegenwart06}
Note, however, that the metamagnetism of the latter system is in vicinity to a tricritical point associated with antiferromagnetic ordering,\cite{Aoki11-2,Aoki12} and alternative interpretations have been also proposed.\cite{Daou06,Pfau12,Bercx12}

In fact, the presence of metamagnetic QCEPs might be generic to itinerant ferromagnets. It has been suggested that the ferromagnet transition in metals generally becomes first-order upon tuning the Curie temperature towards zero.\cite{Belitz05} The resulting tricritical point gives rise to surfaces of first-order metamagnetic transitions at finite $H$ that terminate in a QCEP. There are indications that such a scenario applies to a number of ferromagnets, for example, ZrZn$_2$ (Ref.~\onlinecite{Uhlarz04}) and UGe$_2$ (Ref.~\onlinecite{Taufour10}). In a recent study, Aoki {\it et al.}\cite{Aoki11} demonstrated that UCoAl exhibits a first-order ferromagnetic instability, and its metamagnetic QCEP can be reached by applying a pressure $p=1.5$GPa. 

In a previous publication, we theoretically discussed the universal critical thermodynamic signatures expected close to a metamagnetic QCEP.\cite{Weickert10} The metamagnetic instability is defined by a diverging differential susceptibility $\chi(T)$ at a finite $H_m$, which together with the 
Ising symmetry of the endpoint implies a number of characteristic phenomena. In the presence of a magnetoelastic coupling, one finds a diverging Gr\"uneisen parameter, a sign change of thermal expansion as a function of field, $\alpha(H)$, and a minimum in the specific-heat coefficient $\gamma(H)$. Moreover, the susceptibility, magnetostriction, and compressibility should exhibit the same metamagnetic singularity and, as a result, one expect a pronounced crystal softening. 

In this work, we illustrate and compute  these signatures within a specific model, i.e., the spin-fluctuation theory for quantum critical metamagnetism, which was put forward by Millis {\it et al.}\cite{Millis02,Schofield02} extending earlier work by Moriya\cite{Moriya86} and Yamada.\cite{Yamada93} We derive the detailed magnetic field and temperature dependence of thermodynamic quantities, specify the crossover lines in the phase diagram, and give details for the predicted behaviour of thermal expansion that already appeared in a short publication.\cite{Gegenwart06}

The outline of the paper is as follows. In section \ref{sec:model} we introduce the model of Millis {\it et al.}\cite{Millis02,Schofield02}, derive the phase diagram and compute the free energy density. In section \ref{sec:thermodynamics} we present the result for thermodynamic quantities, and we conclude in section \ref{sec:summary} with a summary and discussion. 

\section{Spin-fluctuation theory for itinerant metamagnetism}  
\label{sec:model}  

We follow Millis {\it et al.}\cite{Millis02,Schofield02} and consider spin-fluctuations in a metal close to a metamagnetic instability. We 
concentrate on the immediate vicinity of the instability where the longitudinal part of the Pauli magnetization $M$, becomes critical. The fluctuations around its mean value, $\phi = M - \langle M \rangle$, identifies an Ising order parameter that is governed by the potential
\begin{align} \label{BarePotential}
\mathcal{V}_0(\phi) = - h_0 \phi + \frac{r_0}{2!} \phi^2 + 
\frac{u_{30}}{3!} \phi^3 + 
\frac{u_{40}}{4!} \phi^4 + 
\frac{u_{50}}{5!} \phi^5 + \dots
\end{align} 
The potential $\mathcal{V}_0(\phi)$ has no specific symmetries and contains in principle all powers of $\phi$. However, later we will use the freedom to shift the field $\phi$ by a constant in order to eliminate the cubic term from the renormalized effective potential. 
Close to the metamagnetic QCEP, the renormalized parameters in front of the linear and quadratic terms, i.e., $h$ and $r$, are small. The linear term is proportional to the deviation of $H$ from the critical field, $h \propto H-H_m$.

The potential for the Ising order parameter is promoted to a quantum critical theory by allowing for spatial and temporal variations. The resulting action reads\cite{Millis02,Schofield02}
\begin{align} \label{QCEPTheory}
\mathcal{S} &= \int_0^\beta d\tau \int d^d x 
\left[
\frac{1}{2} \phi\, \Delta(-i \partial_x, i \partial_\tau) \phi
+ \mathcal{V}_0(\phi) \right]
\end{align}
with the gradient terms given by
\begin{align} \label{gradients}
\Delta(k,\omega_n) &= k^2 + \frac{|\omega_n|}{k^{z-2}} ,
\end{align}
where $\omega_n$ is a bosonic Matsubara frequency and the dynamical exponent is $z=3$. Similar to ferromagnetic paramagnons, the dynamics of the metamagnetic fluctuations is dominated by Landau damping due to the excitation of particle-hole pairs in the metal. 
In Eq.\eqref{gradients} and in the following, we use dimensionless units for lengths and energies as well as for the Ising order parameter $\phi$.

The spin-fluctuation theory for metamagnetic quantum criticality rests on the assumptions that $(i)$ the Taylor expansion of the bare potential in Eq.~\eqref{BarePotential} exists and $(ii)$ the static susceptibility is analytic in momenta such that indeed $\Delta(k,0) = k^2$. This requires in particular that the electronic density of states is sufficiently smooth as the expansion coefficients of Eq.~\eqref{BarePotential} are directly related to the derivatives of the density of states at the Fermi level. For example, when the Fermi level is exactly located at a van-Hove singularity the effective bosonic theory \eqref{QCEPTheory} is not applicable. In addition, non-analyticities in a critical Fermi liquid can also been induced by interaction effects, which have been intensively investigated in recent years, see e.g.~Refs.~\onlinecite{Belitz97,Rech06,Maslov09}. It is established that for a ferromagnetic instability, i.e., a magnetic instability at zero magnetic field $H=0$, virtual particle-hole pairs can mediate long-range interactions between the fluctuations of the order parameter which in turn result in non-analyticities of the Landau potential that render the transition first-order.\cite{Belitz05}  At a metamagnetic instability at finite field $H_m > 0$, however, spin-flip processes are frozen out at low temperatures due to the finite Pauli magnetization, and the Ising order parameter only couples to the relative density fluctuations of the spin-majority  and -minority Fermi surfaces. It was argued in Ref.~\onlinecite{Rech06} that in this case the interaction mediated by particle-hole pairs is less singular, and the spin-fluctuation theory remains analytic. Consequently, it is believed that the assumptions $(i)$ and $(ii)$ hold and that the theory \eqref{QCEPTheory} is well defined.

\subsection{Fluctuation corrections to the effective potential}

The fluctuations of the field $\phi$ will dominate the low-temperature thermodynamics close to the metamagnetic QCEP. 
These fluctuations can be treated in a perturbative manner. It is nevertheless convenient to apply the method of the renormalization group (RG) to incorporate these corrections in a systematic self-consistent fashion.\cite{Millis02} For the problem of a QCEP, we find it convenient to apply the functional renormalization group (RG) and consider the RG flow of the effective potential $\mathcal{V}$ itself.
 
The potential $\mathcal{V}$ at some RG scale $\lambda$ can be expressed as a line integral along an RG trajectory \cite{Wegner73,Berges02}
\begin{align} \label{RGFlow}
\mathcal{V}_\lambda(\phi)
&= \mathcal{V}_0(\phi) + \int_0^\lambda d\lambda' e^{-D \lambda'} 
f(T e^{z \lambda'}, \mathcal{V}''_{\lambda'}(\phi) e^{2\lambda'}),
\end{align}
where $\mathcal{V}''$ is the second derivative of the potential with respect to the field $\phi$.
The effective dimensionality is $D=d+z > 4$ with the spatial dimension $d=2$ or $d=3$, and $z=3$. The 
function $f$ is given by
\begin{align} \label{fFunction}
f(T,R) = \Lambda \frac{\partial}{\partial \Lambda} 
\frac{1}{2} \int \frac{d^d k}{(2\pi)^d} \frac{1}{\beta} \sum_{\omega_n} 
\log \left[R + \Delta(k,\omega_n) \right].
\end{align}
The momentum integral and the sum over bosonic Matsubara frequencies is here understood to be regularized with some UV cutoff $\Lambda$. Replacing the Matsubara sum with an integral along the branch cut singularity and employing a hard cutoff regularization, one obtains
\begin{align} 
f(T,R) = - \frac{K_d}{2\pi} \Lambda \frac{\partial}{\partial \Lambda} 
 \int^{\Lambda^z}_0 d\epsilon \int_0^\Lambda dk\, k^{d-1} 
\coth\frac{\epsilon}{2 T} \\\nn
\times \arctan \frac{\epsilon\, k^{2-z}}{R+k^2}
\end{align}
where $K_d = (2^{d-1} \pi^{d/2} \Gamma(d/2))^{-1}$. 

The effective potential is obtained after incorporating all fluctuation corrections by integrating over the full RG trajectory, $\mathcal{V} \equiv \mathcal{V}_\infty$. Two different types of fluctuation corrections can be distinguished corresponding to the separation of the $f$ function into two parts, 
\begin{align}
f(T, R) = f_0(R) + f_\infty(T, R),
\end{align}
with
\begin{align}
f_0(R) = f(0,0) + f^{(0,1)}(0,0) R + \frac{1}{2} f^{(0,2)}(0,0) R^2,
\end{align}
where $f^{(0,n)}$ is the n$^{\rm th}$ derivative with respect to the second argument. The correction arising from $f_0$ contributes considerably only at the initial stage of the RG flow, i.e., for small values of $\lambda$ in Eq.~(\ref{RGFlow}). These corrections can be conveniently absorbed into a renormalization of the bare parameters of the potential (\ref{BarePotential}),
\begin{align} \label{RenPotential}
\mathcal{V}_{\rm ren}(\phi) &\equiv
\mathcal{V}_0(\phi) + \int_0^\infty d\lambda\, e^{-D \lambda} 
f_0(\mathcal{V}''_{0}(\phi) e^{2\lambda})
\nn\\
&= - h \phi + \frac{r}{2!} \phi^2 +  
\frac{u}{4!} \phi^4 + 
\frac{u_{5}}{5!} \phi^5 + \dots
\end{align} 
In the last line we disregarded an uninteresting constant. As announced, the freedom to shift the field $\phi$ allows to eliminate the cubic coupling term. 

On the other hand, the contribution of the fluctuations attributed to $f_\infty$ only develops at the final stage of the RG trajectory, i.e., for large $\lambda$ and comprises the universal fluctuation corrections. So we finally obtain for the effective potential
\begin{align} \label{EffPotential}
\mathcal{V}(\phi) &= 
\mathcal{V}_{\rm ren}(\phi)
+ T^{D/z} \mathcal{A}_d(\mathcal{V}''(\phi) T^{-2/z}) ,
\end{align}
where we introduced the function 
\begin{align} \label{functionA}
\mathcal{A}_d(x) = \int^\infty_{-\log \Lambda/T^{1/z}} d \mu\, e^{-D \mu} \Lambda^{-D} f_\infty (\Lambda^z e^{z\mu}, \Lambda^2 e^{2\mu} x).
\end{align} 
The effective potential is determined by a differential equation (\ref{EffPotential}). However, it turns out that this differential equation can be easily solved perturbatively for small temperatures $T$. The expression (\ref{EffPotential}) for the effective potential is the fundamental equation of this paper. 
 
We will need the limits of $\mathcal{A}_d$ for small and large arguments. For dimension $d=2,3$ the limiting behaviour can be summarized as follows
\begin{align} \label{LimitsOfA}
\lefteqn{\mathcal{A}_d(x) =}\\\nn
&= 
\left\{
\begin{array}{ll}
- \mathfrak{a}_1 + \mathfrak{a}_2 x - \mathfrak{a}_3 x^{\frac{d}{2}} + \dots
& x\ll1
\\
x^{\frac{D}{2}} \left(\bar{\mathfrak{a}}_1 - \bar{\mathfrak{a}}_2\, x^{-z} + \bar{\mathfrak{a}}_3\, x^{- \frac{z(d-z+4)}{2}} +
\dots \right)
& x\gg 1
\end{array}
\right.
\end{align}
where the coefficients might depend logarithmically on $x$.
The non-analytic term with coefficient $\mathfrak{a}_3$ originates from the zero Matsubara mode. 
The term with coefficient $\bar{\mathfrak{a}}_3$ will give rise to non-analytic Fermi liquid corrections. The explicit form of the coefficients are given in Appendix \ref{app:Details}.

We will limit ourselves to situations where all parameters of the potential $\mathcal{V}_{\rm ren}$ 
are positive except the tuning parameter $h$, which can have either sign. We will therefore not discuss first-order metamagnetic transitions, which are realized for $r<0$. In the limit $r = 0$ one obtains the QCEP. The quartic, $u$, quintic, $u_5$, and higher order couplings are formally irrelevant in the RG sense as the model is above its upper critical dimension, $D>4$. The fourth order term is however needed in order to stabilize the potential. The influence of the quintic term is negligible at lowest temperatures close to the QCEP and can be omitted, see Appendix \ref{app:QuinticTerm}. The most relevant coupling is $h$. It is the tuning parameter of the metamagnetic transition $h \propto H - H_m$, where $H_m$ is the critical magnetic field. After disregarding the quintic and higher order terms an Ising symmetry of the potential emerges with respect to the transformation $h \to -h$. This implies that the critical free energy density will be an even function of $h$, $\mathcal{F}_{\rm cr}(-h) = \mathcal{F}_{\rm cr}(h)$, in the limit of lowest temperatures. 
 
The critical free energy density $\mathcal{F}_{\rm cr}$ is obtained by taking the effective potential at its minimum value, $\mathcal{F}_{\rm cr} = \mathcal{V}(\bar{\phi})$. In the following the solution $\bar{\phi}$ will be discussed.
  
\subsection{Classical field configuration}

The classical field configuration minimizes the effective potential, $\mathcal{V}'(\bar{\phi}) = 0$. The tuning parameter $h$ couples linearly to the field $\phi$ and thus acts as a force. The field within the potential adjusts to this force so that the system is in equilibrium. The restoring force 
which counterbalances $h$ depends on the strength of the potential. In the following we distinguish a linear and a non-linear regime for the stiff and soft potential, respectively.

\begin{figure}
\includegraphics[width= 0.95\linewidth]{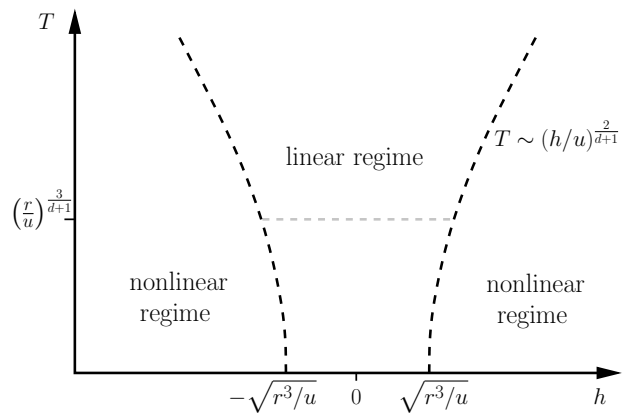}
\caption{\label{fig:ResponseRegimes} 
The linear regime, $R^3 \gg u h^2$, is centered around $h=0$. The crossover lines are bent outwards from the center. At large temperatures the boundary follows an asymptotic powerlaw in $h$. At a temperature $T \sim (r/u)^{3/(d+1)}$ the boundaries start to repel and hit the zero temperature axis with an infinite slope.
}
\end{figure}

\subsubsection{Linear regime $R^3 \gg u h^2$}
 
In the linear regime, the effective potential is simply obtained by expanding (\ref{EffPotential}) in the field  $\phi$. Retaining only the leading temperature correction we obtain
\begin{align} \label{EffPotExpanded}
\mathcal{V}(\phi) \approx \mathcal{V}(0) - h \phi + \frac{R}{2!} \phi^2 + \frac{u}{4!} \phi^4 .
\end{align}
For a short discussion of sub-leading temperature corrections see Appendix \ref{app:QuinticTerm}. In the linear regime, $R^3 \gg u h^2$, the quadratic dominates over the quartic term and the classical field configuration is given by
\begin{align} \label{ClassicalField-Linear}
\bar{\phi} = \frac{h}{R}, \quad\quad R^3 \gg u h^2.
\end{align}
The stiffness of the potential $R$ is defined by the equation
\begin{align} \label{EffMass}
R
&= r +  u\, T^{(D-2)/z} \mathcal{A}'_d(R\, T^{-2/z}) 
\end{align}
with the first derivative of the function $\mathcal{A}_d$ defined in (\ref{functionA}). Keeping only the leading temperature correction, $R$ is approximately given by
\begin{align} \label{EffMassLimits}
R
&=  r + u
\left\{
\begin{array}{ll}
\mathfrak{r}_1\, T^{\frac{d+1}{3}} 
& 
\quad R \ll T^{2/3}
\\
\mathfrak{r}_2\, T^2 r^\frac{d-5}{2} & 
\quad
R \gg T^{2/3}
\end{array}
\right.
\end{align}
with positive coefficients $\mathfrak{r}_i$, and we set $z=3$ explicitly. Note that a finite temperature enhances the stiffness. For small arguments of $\mathcal{A}'_d$, the stiffness, $R$, obtains a temperature dependence, $T^{(D-2)/z}$. 
In the opposite limit, $R \ll T^{2/z}$, the thermal fluctuations lead to a temperature correction of Fermi liquid type, $T^2$. 

Taking the effective potential at the minimum, $\mathcal{V}(\bar{\phi})$, we obtain the free energy density. In lowest order in the tuning parameter $h$ we get 
\begin{align} 
\mathcal{F}_{\rm cr} = T^{D/z} \mathcal{A}_d(R\, T^{-2/z}) - \frac{h^2}{2 R},
\quad {\rm for}\quad R^3 \gg u h^2.
\end{align}
The first term is due to the Gaussian fluctuations around the local minimum; the second term is the energy stored in the potential $\mathcal{V}$, and it is quadratic in $h$ in the linear regime. Note that this latter contribution is temperature dependent due to fluctuation corrections to the potential's stiffness, $R$.
 
\subsubsection{Non-linear regime $R^3 \ll u h^2$}

In the non-linear regime $R^3 \ll u h^2$ the potential (\ref{EffPotential}) is soft and its curvature is determined by the field itself, $\mathcal{V}''(\phi) \approx \frac{u}{2} \phi^2$. So we get
\begin{align}
\mathcal{V}(\phi) &\approx -h \phi 
+ \frac{u}{4!} \phi^4
+ T^{D/z} \mathcal{A}_d\left(\frac{u}{2} \phi^2 T^{-2/z}\right).
\end{align}
The restoring force is now determined by the quartic term. In lowest order in temperature, the potential is minimized by 
\begin{align} \label{ClassicalField-Nonlinear}
\bar{\phi} =  \left|\frac{6 h}{u}\right|^{1/\delta} {\rm sign}(h), \quad\quad R^3 \ll h^2 u
\end{align}
with mean-field exponent, $\delta=3$. The resulting free energy reads
\begin{align} \label{FreeEnergy-Nonlinear}
\mathcal{F}_{\rm cr} = T^{D/z} \mathcal{A}_d\left( 
\frac{6^{2/3} u^{1/3} |h|^{2/3}}{2 T^{2/z}}\right) - \frac{3^{4/3}}{2^{5/3}} \frac{|h|^{4/3}}{u^{1/3}}. 
\end{align}
The term proportional to $|h|^{4/3}$ reflects the non-linear nature of the restoring force, and it is temperature independent in the non-linear regime. This scaling only applies as long as $|h| \ll h_\Lambda$ where the cutoff-field $h_\Lambda$ is determined by the quintic coupling, see Appendix \ref{app:QuinticTerm}. Moreover, the corrections to the free energy arising from the temperature correction to the classical field configuration (\ref{ClassicalField-Nonlinear}) is sub-leading, see Appendix \ref{app:FieldCorrection}.

The crossover boundaries in the $(h,T)$ plane between the linear and non-linear regimes are shown in Fig.~\ref{fig:ResponseRegimes}. 
In the limit $r \ll u\, T^{(D-2)/z}$, when $R$  is dominated by the thermal fluctuations the boundaries follow $T \sim (h/u)^{2/(d+1)}$. At a temperature $T \sim (r/u)^{3/(d+1)}$ the boundaries start to repel each other and hit the zero temperature axis with an infinite slope, $h \sim \pm \sqrt{r^3/u} + \mathcal{O}(T^2)$.
 
\subsection{Free energy density}
\label{subsec:FreeEnergy}

\begin{figure}
\includegraphics[width= 0.9\linewidth]{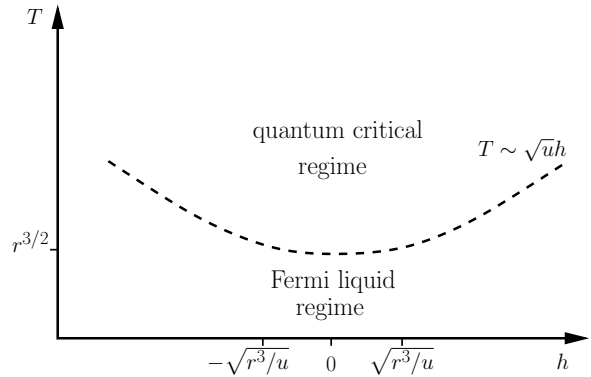}
\caption{\label{fig:FL-NFL-crossover} 
The local curvature near the potential minimum, $T \sim (\mathcal{V}''(\bar{\phi}))^{z/2}$, defines a crossover scale between a quantum critical and a Fermi liquid regime. The QCEP obtains in the limit $r\to 0$ when the quantum critical regime touches the zero temperature axis at $h=0$. For large $|h| \gg \sqrt{r^3/u}$ the crossover boundary increases linearly with $h$.
}
\end{figure}

The free energy is determined by the effective potential at the position of its minimum, $\mathcal{F}_{\rm cr} = \mathcal{V}(\bar{\phi})$. The argument of the function $\mathcal{A}_d$ in
the expression for the effective potential (\ref{EffPotential}) identifies a second important crossover scale in the phase diagram, $\mathcal{V}''(\bar{\phi}) \sim T^{2/z}$. The local curvature is approximately given by $\mathcal{V}''(\bar{\phi}) \approx R + \frac{u}{2} \bar{\phi}^2$ with the stiffness $R$ of the potential, see Eq.~(\ref{EffMass}).
For large local curvature, $\mathcal{V}''(\bar{\phi}) \gg T^{2/z}$, the free energy has a standard Fermi liquid form, $\mathcal{F}_{\rm cr} = \mathcal{F}_0 - (T/T_0)^2$, and thermodynamics is conventional for a metal. In the limit of small curvature, $\mathcal{V}''(\bar{\phi}) \ll T^{2/z}$, thermal fluctuations induce non-analytic $T$ dependences resulting in unusual thermodynamic behaviour. The crossover line between this Fermi liquid and quantum critical regime is shown in Fig.~\ref{fig:FL-NFL-crossover}.

\begin{figure}
\includegraphics[width= 0.9\linewidth]{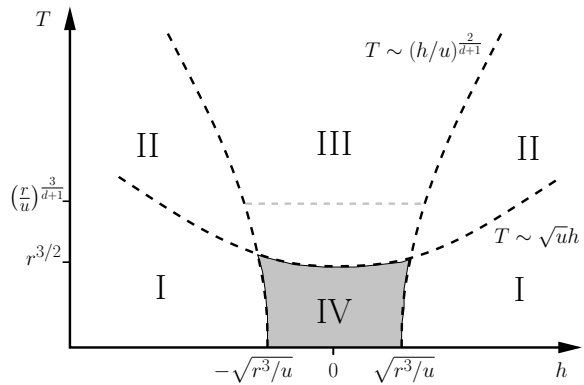}
\caption{\label{fig:regimes} 
Different regimes resulting from the two crossover lines of Fig.~\ref{fig:ResponseRegimes}
and \ref{fig:FL-NFL-crossover}. In the limit $r\to 0$ the regime IV shrinks to a point coinciding with the QCEP. At finite $r$ all critical divergencies are cut off upon entering regime IV.
}
\end{figure}

The intersection between the crossover lines of Fig.~\ref{fig:ResponseRegimes} and \ref{fig:FL-NFL-crossover} define four different regimes, see Fig.~\ref{fig:regimes}. Note that within regime III there is an additional crossover at $r \sim u T^{(D-2)/z}$ depending on whether the stiffness $R$, (\ref{EffMass}), is dominated by $r$ or the thermal fluctuations. 
For dimensions $d=2,3$ the leading behaviour of the free energy in the four regimes can be summarized as follows,
\begin{align} \label{FreeEnergyLimits}
\mathcal{F}_{\rm cr} =
\left\{
\begin{array}{cc}
\displaystyle
- \mathfrak{f}_1\, T^{2} (u h^{2})^{\frac{d-3}{6}} 
 - \mathfrak{f}_2\, \frac{|h|^{4/3}}{u^{1/3}}
& \quad{\rm I}\\[1em]
\displaystyle
-\mathfrak{f}_3\, T^{\frac{d}{3}+1} 
+ \mathfrak{f}_4\, T^{\frac{d+1}{3}} (u h^2)^{1/3} 
- \mathfrak{f}_2\, \frac{|h|^{4/3}}{u^{1/3}}
& \quad{\rm II}\\[1em]
\displaystyle
-\mathfrak{f}_5\, T^{\frac{d}{3}+1}  
- \frac{h^2}{2 R}
& \quad{\rm III}\\[1em]
\displaystyle
- \mathfrak{f}_6\, T^2 r^{\frac{d-3}{2}}
- \frac{h^2}{2 R}
& \quad{\rm IV}
\end{array}
\right.
\end{align}
where we set explicitely $z=3$. The coefficients $\mathfrak{f}_i$ are all positive and depend only logarithmically on the external parameters, see Appendix~\ref{app:Details}. It is important to realize that $R$ is temperature dependent (\ref{EffMassLimits}). It is in fact this temperature dependence which leads to some strong thermodynamic signatures in the vicinity of the metamagnetic transition.

\section{Thermodynamics of the quantum critical metamagnetic crossover}
\label{sec:thermodynamics}

Thermodynamics measures the response of the free energy 
upon variation of external control parameters like temperature, $T$, pressure, $p$, or magnetic field, $H$. There are several interesting second order derivatives of the free energy whose behavior is discussed in the following. The specific heat coefficient quantifies the change of the free energy upon temperature variations and is defined as $\gamma = - \frac{\partial^2 \mathcal{F}}{\partial T^2}$. The thermal expansion is a mixed derivative with respect to temperature and pressure, $\alpha = \frac{1}{V} \frac{\partial V}{\partial T}|_{p} = \frac{1}{V}\frac{\partial^2 \mathcal{F}}{\partial T\partial p}$. The compressibility is given by $\kappa = - \frac{1}{V} \frac{\partial V}{\partial p}|_{T} = - \frac{1}{V}\frac{\partial^2 \mathcal{F}}{\partial p^2}$, the magnetostriction is a mixed derivative with respect to pressure and magnetic field, $\lambda = \frac{1}{V}\frac{\partial V}{\partial H} = - \frac{1}{V}\frac{\partial^2 \mathcal{F}}{\partial p \partial H}$ and the magnetic susceptibility is $\chi = -\frac{\partial^2 \mathcal{F}}{\partial H^2}$. We will also discuss the Gr\"uneisen ratio $\Gamma = \alpha/C$ and its magnetic analogue $\Gamma_H = -\frac{\partial M}{\partial T}/C$ where $C = \gamma T$.\cite{Zhu03}

\subsection{Dependencies of the scaling fields}

In order to determine thermodynamics of metamagnetism, we have to discuss the dependence of the free energy (\ref{FreeEnergyLimits}) on pressure, magnetic field and temperature. The latter plays a special role as it appears explicitly in the effective theory via the dynamics of the metamagnetic order parameter, see Eq.~(\ref{gradients}), resulting in an explicit $T$-dependence of the free energy (\ref{FreeEnergyLimits}) that is responsible for the leading low-$T$ behavior. Besides this explicit $T$-dependence, there is also an implicit dependence via the scaling fields, i.e., the coefficients of the effective potential, see Eq.~(\ref{RenPotential}). They also depend on pressure and magnetic field. We have to consider the implicit dependence of the linear and the quadratic coefficient of the potential, $h$ and $r$, respectively,
\begin{align} \label{IntrinsicDependences}
h &= h(H,p,T) \approx h(H,p)  + h_T T^2
\\
r &= r(H,p,T) \approx r(H,p). 
\end{align}   
In principle, their dependence can be obtained from microscopic calculations.\cite{Binz04} Within our effective theory, we assume that the pressure and magnetic field dependence is sufficiently weak such that the derivatives of $h$ and $r$ with respect to $p$ or $H$ can be treated as a constant near the metamagnetic transition, e.g.~$\partial_p h \approx$ const.. In addition, the intrinsic temperature dependence is assumed to be of Fermi liquid type, $T^2$. The temperature dependence of the quadratic coefficient, $r$, can be neglected close to the QCEP. However, as we will see, the $T^2$ dependence of the linear coefficient with amplitude $h_T$ leads to important corrections although $h_T$ is formally irrelevant in the RG sense.

The following discussion is divided into two parts. In the first part the implicit $p$ and $H$ dependence of only the most relevant parameter, $h(H,p)$, is considered and $h_T$ and the dependences of $r$ are neglected. In this case, the emergent Ising symmetry of the potential (\ref{RenPotential}) will be directly reflected in the physical phase diagram spanned by temperature, $T$, and magnetic field, $H$. In a second step, we will consider the corrections induced by $h_T$ and the dependences of $r(p,H)$.
 
\subsection{Emergent Ising symmetry in thermodynamics}

Taking into account only the $p$ and $H$ dependence of the control parameter, $h(H,p) \propto H - H_m(p)$, the critical free energy is of the form,
\begin{align} \label{reducedFreeEnergy}
\mathcal{F}_{\rm cr}(H,p,T) \approx \mathcal{F}_{\rm cr}(H-H_m(p), T).
\end{align}
Near the metamagnetic transition the control parameter measures the distance to the critical field, $h \propto H - H_m(p)$, where the critical field is assumed to depend smoothly on pressure, $\partial_p H_m \approx$ const.. With these assumptions it is clear that the derivative of the free energy with respect to magnetic field or pressure yield the same thermodynamic information. As a consequence,
the second order derivatives of the free energy are related,
\begin{align} \label{DerivativesOfF}
\frac{\partial^2 \mathcal{F}_{\rm cr}}{\partial T^2} &\propto \gamma;
\quad
\frac{\partial^2 \mathcal{F}_{\rm cr}}{\partial h \partial T} 
\propto \alpha, \frac{\partial M}{\partial T}; 
\quad
\frac{\partial^2 \mathcal{F}_{\rm cr}}{\partial h^2} \propto \chi, \lambda, \kappa.
\end{align}
For example, the susceptibility $\chi$ is asymptotically proportional to the critical part of the compressibility $\kappa$. 
In fact, their proportionality can be taken as a criterion for the assumption (\ref{reducedFreeEnergy}) to be valid. Note that the free energy $\mathcal{F}_{\rm cr}$ only accounts for the critical metamagnetic contributions. There might exist non-critical background contributions as, for example, for the compressibility.   

Another immediate consequence of the form (\ref{reducedFreeEnergy}) for the free energy is that the emergent Ising symmetry of the 
potential (\ref{RenPotential}) results in a reflection symmetry with respect to the metamagnetic field $H_m$, 
\begin{align} \label{FreeEnergyIsingSymm}
\mathcal{F}_{\rm cr}(H-H_m(p),T) = \mathcal{F}_{\rm cr}(H_m(p)-H,T).
\end{align}
The asymptotic Ising symmetry of the problem thus becomes explicit in the thermodynamic signatures. This symmetry implies for example that the thermal expansion as a function of $H$ changes its sign at the critical field $H_m$, and that specific heat and susceptibility are even functions of $H-H_m$. Deviations from this symmetry is an important indicator for additional contributions to the reduced form of $\mathcal{F}_{\rm cr}$, (\ref{reducedFreeEnergy}).

In the following, the magnetization and the behavior of the second order derivatives (\ref{DerivativesOfF}) in the four regimes of Fig.~\ref{fig:regimes} are discussed in detail for general dimensions $d=2,3$ using the expression for the free energy given in Section \ref{subsec:FreeEnergy}, which is correct up to logarithmic corrections. 
Readers who are interested in these logarithmic corrections should consult Appendix \ref{app:Details}.

\subsubsection{Magnetization}

The hallmark of metamagnetism is the strong increase of magnetization, $M$, as a function of magnetic field. From the free energy (\ref{FreeEnergyLimits}) 
we obtain a mean-field behaviour for the change in the magnetization, $\delta M = M - M_m$, see the inset of  
Fig.~\ref{fig:Susceptibility-T}(a). Close to the metamagnetic field in the linear regime of Fig.~\ref{fig:ResponseRegimes}, the magnetization increases linear as a function of $H$ with a susceptibility determined by the stiffness $R$, see Eq.~(\ref{EffMass}). For larger distances, $h \propto H-H_m$, the magnetization crosses over into the non-linear regime and follows $\delta M \sim |h|^{1/\delta}$ with a mean-field exponent $\delta = 3$. The Ising symmetry is reflected in a point symmetry of the change in  magnetization $\delta M$ as a function of $h$. 

\subsubsection{Susceptibility, magnetostriction and compressibility}

\begin{figure}
\includegraphics[width= 0.95\linewidth]{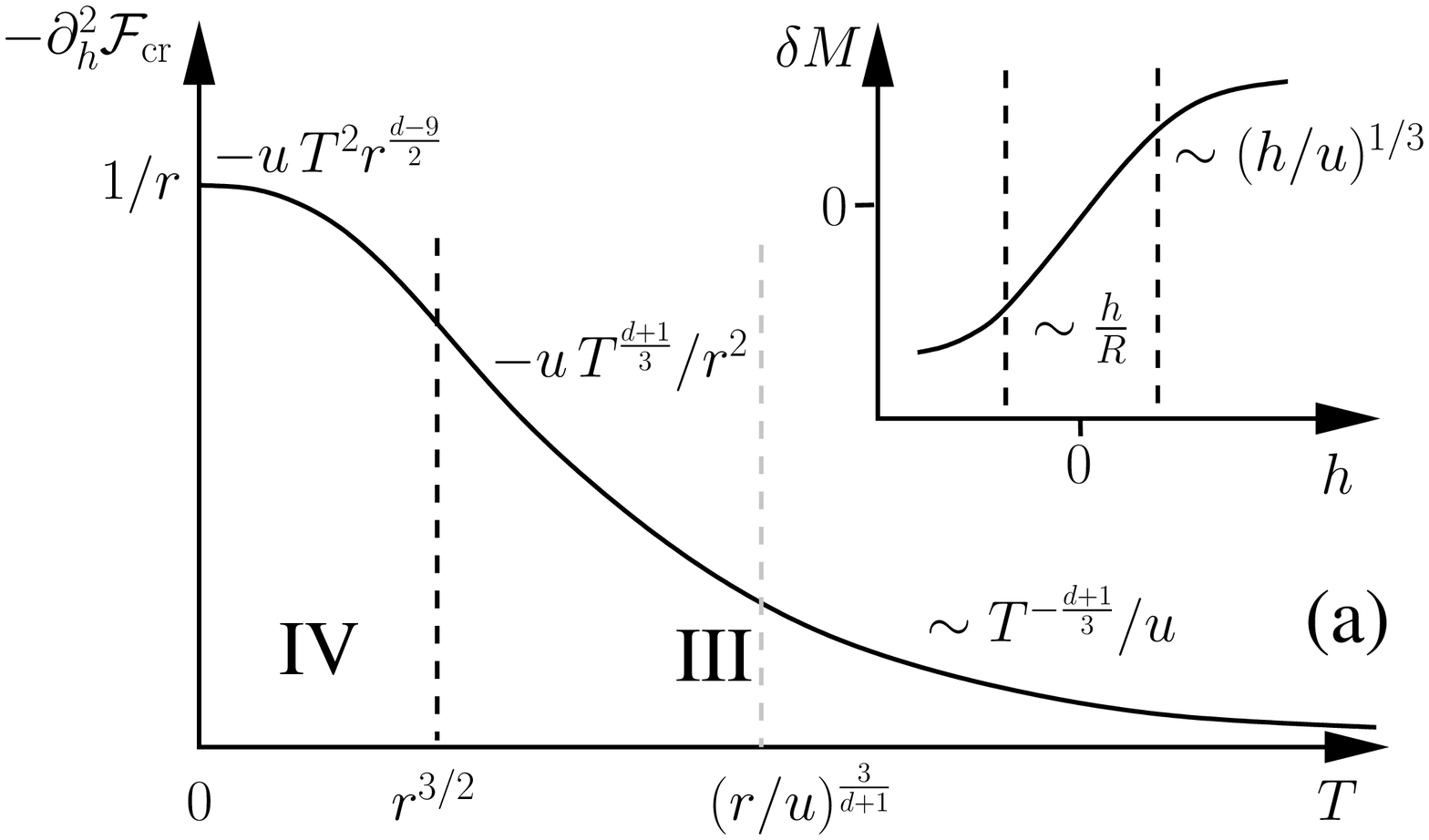}
\includegraphics[width= 0.95\linewidth]{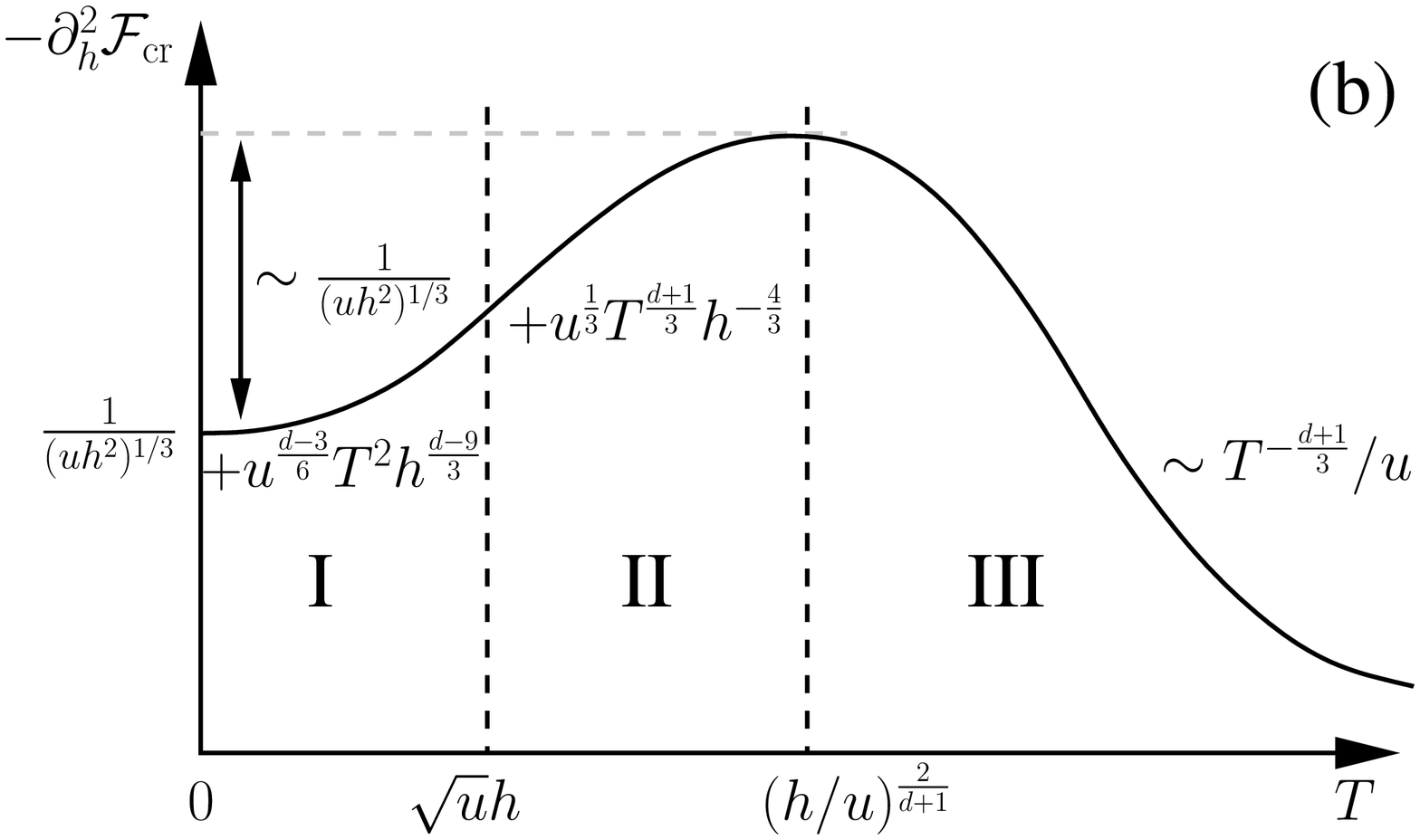}
\caption{\label{fig:Susceptibility-T} 
Sketch of the temperature dependence of the susceptibility $-\partial_h^2 \mathcal{F}_{\rm cr}$, that is an even function of $h$, for (a) fields very close to the metamagnetic transition, $|h| < \sqrt{r^3/u}$, and (b) larger fields such that the non-linear regime is entered for low $T$ (assuming $h>0$), see Fig.~\ref{fig:regimes}. 
Magnetostriction and compressibility are expected to follow the same behaviour. Inset: behaviour of the magnetization change $\delta M = M - M_m$; the crossover lines separate the linear from the non-linear regime, see Fig.~\ref{fig:ResponseRegimes}.
}
\end{figure}
  
Magnetic susceptibility, magnetostriction and compressibility asymptotically follow the same behaviour near the metamagnetic field and are proportional to $-\partial^2_h \mathcal{F}_{\rm cr}$. They are even function of the control parameter, e.g.~$\chi(-h) = \chi(h)$. In Fig.~\ref{fig:Susceptibility-T} a sketch of the temperature dependence is shown. The temperature trace in Fig.~\ref{fig:Susceptibility-T}(a) is located within the linear regime, $|h| < \sqrt{r^3/u}$, see Fig.~\ref{fig:regimes}, and here the susceptibility increases monotonously with decreasing temperature. In this regime the susceptibility is simply given by the inverse stiffness of the potential, $R^{-1}$, see Eq.~(\ref{EffMassLimits}). In Fig.~\ref{fig:Susceptibility-T}(b) a temperature trace for larger $h$ is shown that crosses the boundary to the non-linear regime. At the crossover between III and II the behaviour changes qualitatively leading to a characteristic peak in the susceptibility as already discussed in Ref.~\onlinecite{Millis02}. The difference between the height of the peak and the saturation value at $T=0$ increases as the metamagnetic field is approached.

As the QCEP is approached, the derivative $-\partial^2_h \mathcal{F}_{\rm cr}$ diverges. Note that this implies a diverging negative correction to the compressibility! This indicates that the QCEP is inherently unstable in the presence of a magnetoelastic coupling.\cite{Anfuso08,Garst09}

\subsubsection{Thermal expansion, temperature derivative of magnetization}

\begin{figure}
\includegraphics[width= 0.82\linewidth]{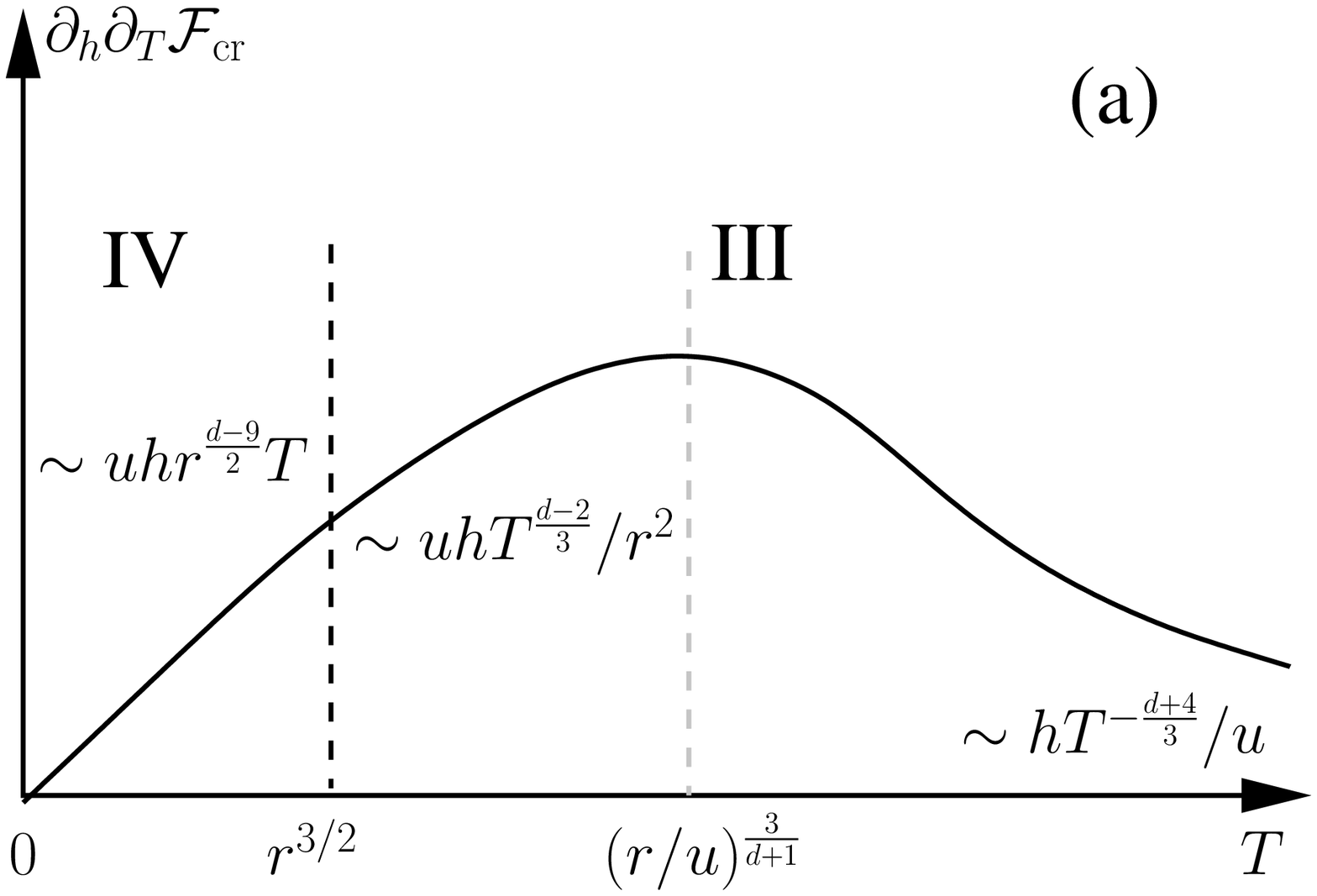}
\includegraphics[width= 0.82\linewidth]{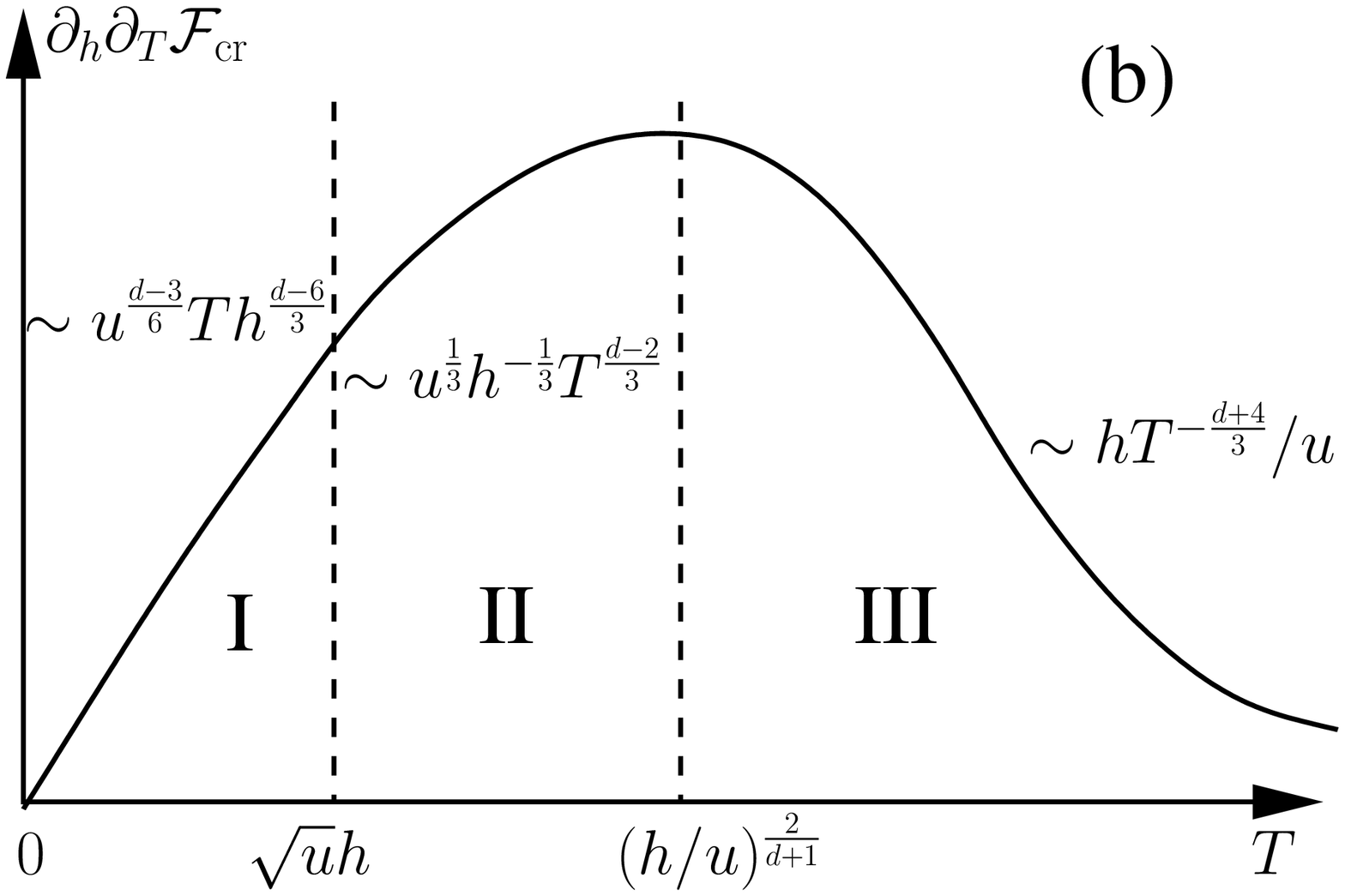}
\caption{\label{fig:ThermalExpansion-T}  Sketch of the temperature dependence of the derivative $\partial_h\partial_T \mathcal{F}_{\rm cr}$ that is proportional to the thermal expansion $\alpha$. It is an odd function of $h$, $\alpha(-h) = - \alpha(h)$, and has a characteristic maximum.
(a) Behaviour very close to the metamagnetic transition $|h| < \sqrt{r^3/u}$ and (b) for larger $h >0$.}
\end{figure}

We discuss the mixed derivative $\partial_h\partial_T \mathcal{F}_{\rm cr}$ representatively in terms of the thermal expansion, $\alpha$. Due to the Ising symmetry, the thermal expansion is an odd function of $h$, $\alpha(-h) = - \alpha(h)$, and changes its sign across the metamagnetic transition. As detailed in Ref.~\onlinecite{Garst05} a sign change of the thermal expansion and, as a consequence, of the Gr\"uneisen parameter, is a generic property of pressure-tuned quantum critical phenomena. The emergent Ising symmetry in the present case implies that the sign change is located exactly at the metamagnetic field $H = H_m$. Corrections to the positions of vanishing thermal expansion are discussed below in Section \ref{sec:IsingViolations}.

A sketch of $\partial_h\partial_T \mathcal{F}_{\rm cr}$ is shown in Fig.~\ref{fig:ThermalExpansion-T}. In the upper panel (a), a temperature trace within the linear regime is shown. Here, the thermal expansion is dominated by the contribution of the condensate, $\partial_h\partial_T \mathcal{F}_{\rm cr} = - h \partial_T R^{-1}$, and is thus proportional to $h$. First, it increases with lowering temperature; at the crossover $T\sim (r/u)^{3/(d+1)}$ within the regime III the thermal corrections cease to dominate the stiffness $R$ of the effective potential resulting in a peak in $\alpha$. Upon entering the Fermi liquid regime IV the thermal expansion vanishes linearly with $T$. In the lower panel (b) the behaviour is shown for $|h| > \sqrt{r^3/u}$ where the temperature trace crosses into the non-linear regime. The maximum follows here the boundary between III and II. Moreover, in regimes I and II the thermal expansion is not simply proportional to $h$ anymore. The behaviour in Fig~\ref{fig:ThermalExpansion-T}(b) was already discussed in Ref.~\onlinecite{Gegenwart06} including a comparison to 
experiments on Sr$_3$Ru$_2$O$_7$.
 
\subsubsection{Specific heat coefficient}

\begin{figure}
\includegraphics[width= 0.95\linewidth]{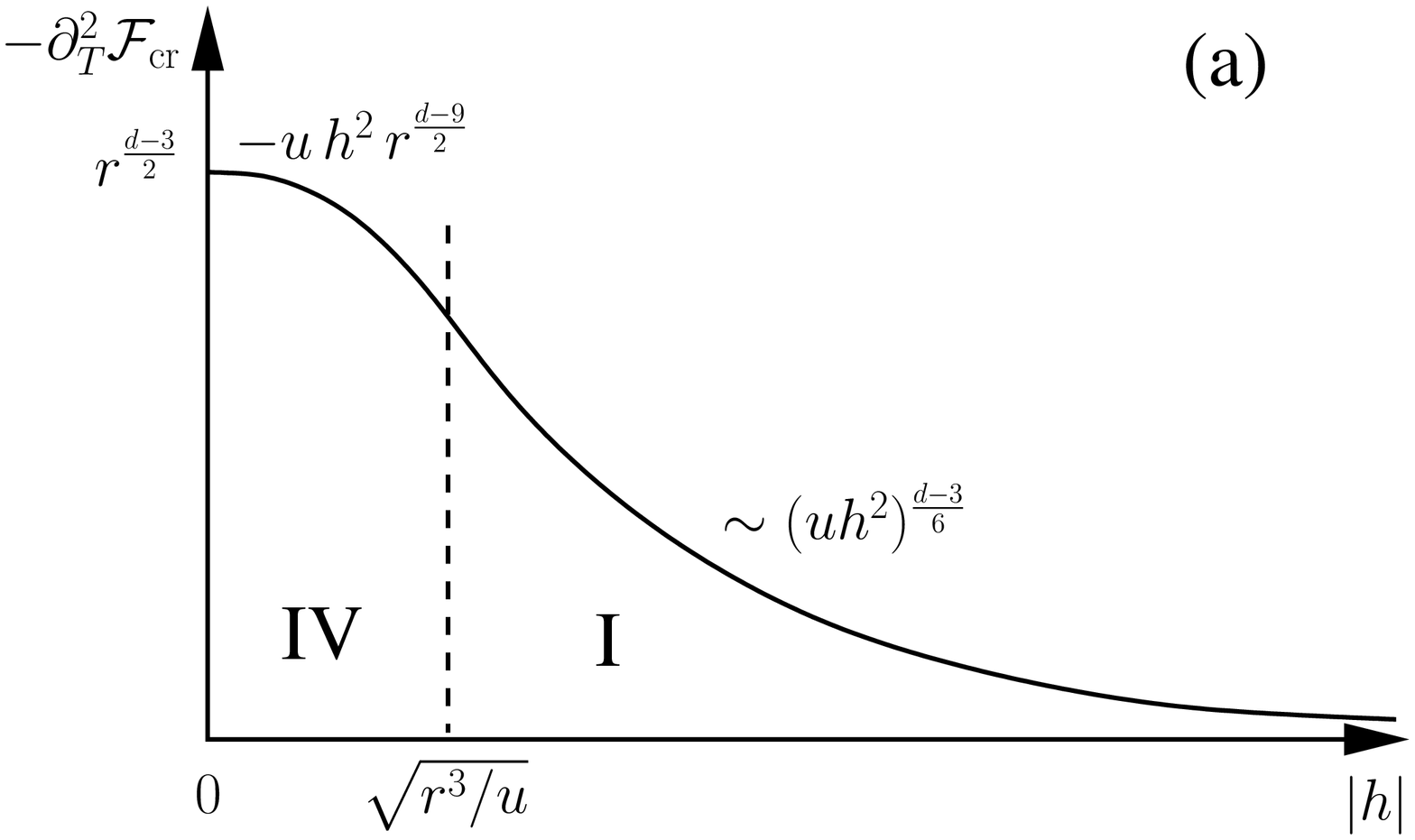}
\includegraphics[width= 0.95\linewidth]{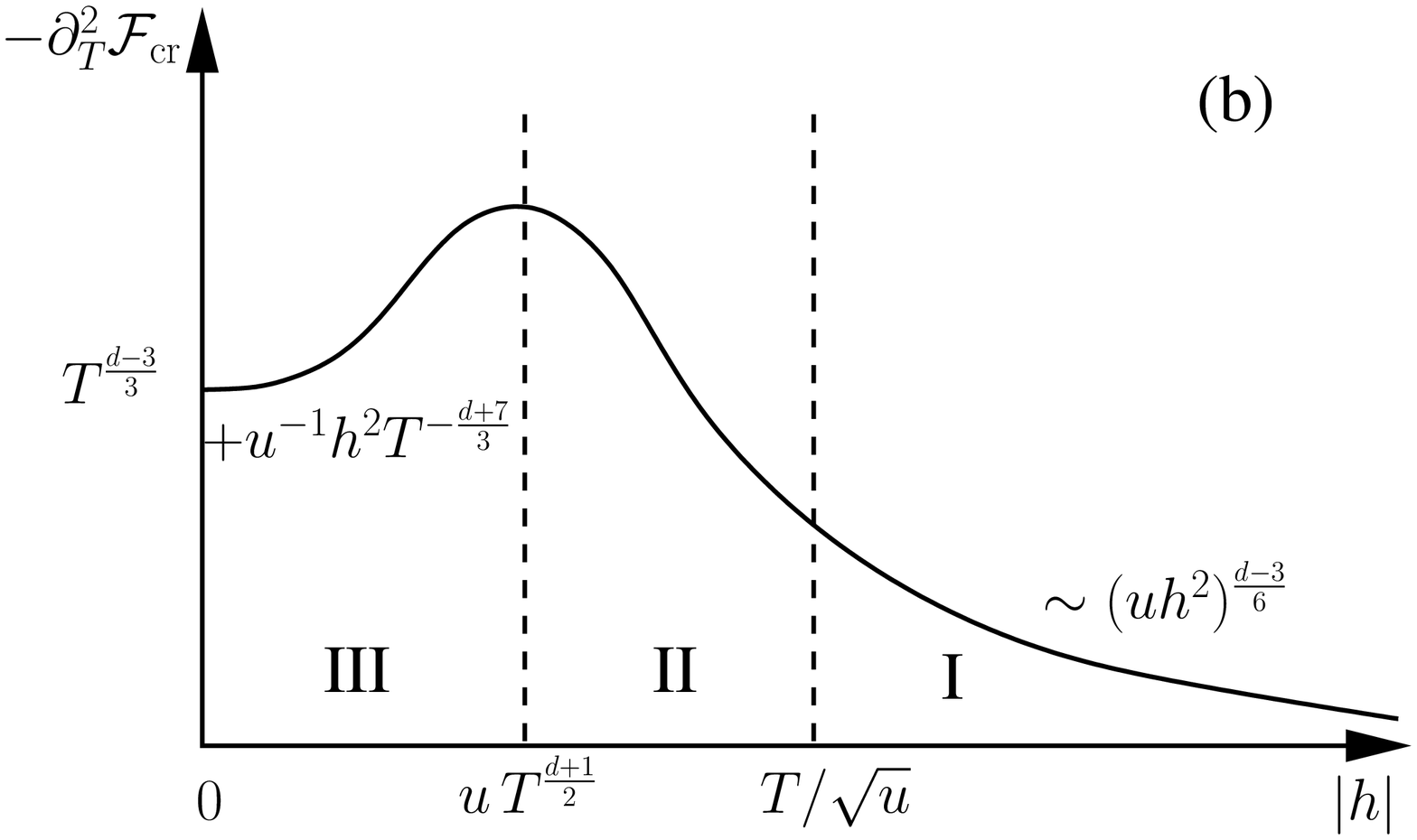}
\caption{\label{fig:SpecificHeat-h} Specific heat coefficient as a function of the tuning parameter $h$, that is an even function of $h$. In panel (a) the behaviour is shown for temperatures within the Fermi liquid regime, see Fig.~\ref{fig:regimes}. Panel (b) displays the behaviour for larger temperatures where the quantum critical regime is entered with a peak at the corresponding crossover.
}
\end{figure}

The specific heat is like the susceptibility an even function of the distance to the metamagnetic transition, $h$. In Fig.~\ref{fig:SpecificHeat-h} the specific heat coefficient, $-\partial_T^2\mathcal{F}_{\rm cr}$, as a function of $h$ is shown. In panel (a) the $h$ trace is located within the Fermi liquid regime, see Fig.~\ref{fig:regimes}. The specific heat coefficient increases monotonously with lowering $h$ and saturates in regime IV at a value determined by $r$. The trace in panel (b) shows the behaviour for larger temperatures as one enters the quantum critical regime. Similar as the susceptibility and the thermal expansion, the specific heat coefficient as a function of $h$ shows a characteristic peak at the boundary between the regimes II and III. The peak is attributed to the contribution of the condensate, $(h^2/2)\partial_T^2 R^{-1}$. For small $h$ and large temperatures it yields a positive correction giving rise to a peak. As a function of temperature this correction changes its sign at $T \sim (r/u)^{3/(d+1)}$ and, as a consequence, for lower temperatures the peak is absent in the $h$ trace. The peak height decreases as $T=h=0$ is approached 
in contrast to the peak height of the susceptibility.

As detailed in Ref~\onlinecite{Weickert10}, such a double peak structure close to the critical field is generally expected whenever the critical susceptibility increases with lowering temperature, $\partial_T^2 \chi(T)>0$. It follows from the Maxwell relation $\partial_H^2 \gamma = \partial_T^2 \chi(T)$ that the curvature of the specific heat coefficient as a function of $H$ coincides with that of the susceptibility as a function of $T$. Consequently, the specific heat coefficient $\gamma(H)$ has a minimum at $H_m$ that is framed by a two peaks.

\subsubsection{Gr\"uneisen parameters}

Within the approximation \eqref{FreeEnergyIsingSymm}, the thermal expansion is proportional to $\partial_T M$ so that the singular parts of the Gr\"uneisen parameter and its magnetic counterpart coincide, $\Gamma \propto \Gamma_H \propto \partial_{h}\partial_T\mathcal{F}_{\rm cr}/(T \partial^2_{T}\mathcal{F}_{\rm cr})$. Due to the Ising symmetry, the Gr\"uneisen parameter changes sign and thus vanishes at the critical field $H_m$ at any finite temperature. In regime III the behavior $\Gamma_H \sim h/(u T^{(4+2d)/3}) $ is obtained.
In regime I the Gr\"uneisen parameter is  simply proportional to the inverse of the tuning parameter, $h \propto H-H_m$,
\begin{align}
\Gamma_H = - \frac{\partial_T M}{C} = \frac{3-d}{3} \frac{1}{H-H_m}
\end{align}
as expected from general scaling arguments.\cite{Zhu03} 
Note, however, that the prefactor does not obey scaling predictions 
because the QCEP for $d=2,3$ is above the upper critical dimension. 
In spatial dimension $d=3$ logarithmic corrections are present, $1/\Gamma_H = (H-H_m) \log\frac{\Lambda^3}{\sqrt{u} |h|} $, see Appendix \ref{app:Details}.

\subsection{Corrections to thermodynamic Ising symmetry: 
locations of entropy accumulation}
\label{sec:IsingViolations}
  
Often, the underlying Ising symmetry of a transition is hidden and not explicit, as for example in the usual liquid-gas transition, because the scaling fields are not simply related to the physical parameters. In the present case, the Ising symmetry is almost explicit close to the metamagnetic field near $T=0$ as the leading temperature dependences are explicitly generated within the theory (\ref{QCEPTheory}). Nevertheless, there are corrections which are especially important whenever the critical leading contribution for some reason vanishes. This is for example the case for the locations of entropy accumulation and thus for the positions of vanishing thermal expansion as discussed in the following.

As mentioned above, the reduced form of the free energy (\ref{FreeEnergyIsingSymm}) implies that the thermal expansion changes its sign exactly at the critical field $H_m$. The locations of vanishing thermal expansion or, equivalently, of the maxima of entropy \cite{Garst05} in the $(H,T)$ plane are thus a very sensitive probe for corrections to the reduced form (\ref{FreeEnergyIsingSymm}). In the following we will study the most important corrections which will shift the position of vanishing thermal expansion away from $h=0$, i.e. $H=H_m$. To this end, we can limit ourselves to the linear regime, i.e., regime III and IV in the phase diagram of Fig.~\ref{fig:regimes}.

The most important corrections derive from the intrinsic temperature dependence of $h$, i.e., the coefficient $h_T$ in the expansion (\ref{IntrinsicDependences}) and the pressure dependence of the quadratic coefficient of the effective potential $r=r(p)$. The latter gives rise to a finite value of the thermal expansion $\alpha$ at $h=0$ that in regime III reads
\begin{align}
\delta \alpha \propto \frac{\partial^2 \mathcal{F}}{\partial r \partial T} \propto T^{\frac{d+z-2}{z}} .
\end{align}
A temperature dependence of this form is familiar from second order quantum phase transitions.\cite{Zhu03} Nevertheless, we will not consider it further here as it is sub-leading compared with the one induced by $h_T$ as we explain in the following.

The thermal expansion within the linear regime is dominated by the condensate contribution to the free energy 
\begin{align}
\alpha \propto \frac{\partial^2 \mathcal{F}}{\partial h \partial T} = - \frac{\partial}{\partial T} \frac{h(T)}{R}, 
\quad R^3 \gg  u h^2,
\end{align}
with $h(T) = h + h_T T^2$. The coefficient $h_T$ gives rise to an {\em additive} correction to thermal expansion. For temperatures $T \gg (r/u)^{3/(d+1)}$ this correction simplifies to 
\begin{align} \label{hTCorrectionLimit}
\delta \alpha \propto \frac{h_T}{u} T^{-\frac{d-2}{3}}.
\end{align}
The two parameters $h_T$ and $u$ are formally irrelevant in the RG sense but they combine to give a singular correction. 
This contribution leads to a shift of the positions of vanishing thermal expansion away from $H=H_m$ thus destroying the explicit, thermodynamic Ising symmetry (\ref{FreeEnergyIsingSymm}). The importance of the intrinsic Fermi liquid temperature dependence for the thermal expansion near the critical field was noticed before in Ref.~\onlinecite{Dzero06}.

In the presence of a finite $h_T$, the locations of vanishing thermal expansion are shifted away from the critical magnetic field, $h \propto  H- H_m = 0$; these locations now follow 
\begin{align}
\left. h \right|_{\alpha = 0} \approx \frac{5-d}{1+d} h_T T^2
\end{align}
 for temperatures $T \gg (r/u)^{3/(d+1)}$.

 \section{Summary \& Discussion}
 \label{sec:summary}
 
We discussed the critical thermodynamics of an itinerant metamagnetic QCEP that is dominated by spin-fluctuations. These fluctuations generate a temperature dependence of the effective potential for the Ising order parameter, that reflects the $z=3$ dynamics of the Landau-damped collective spin-fluctuations. The effective potential was systematically evaluated with the help of the functional renormalization group, and we identified different regimes in the phase diagram, see Fig.~\ref{fig:regimes}. We derived the $H$ and $T$ dependence of the critical free energy density analytically in the different regimes, see Eq.~\eqref{FreeEnergyLimits} and Appendix \ref{app:Details}, and discussed the resulting thermodynamics in section \ref{sec:thermodynamics}. 

In a previous work,\cite{Weickert10} we discussed the universal thermodynamic signatures expected on general grounds close to a metamagnetic QCEP. The present calculations confirm and illustrate these signatures within the spin-fluctuation model. In particular, we find a diverging Gr\"uneisen parameter, a sign change of the thermal expansion $\alpha(H)$, and a minimum in the specific heat coefficient $\gamma(H)$.  
Moreover, the susceptibility, magnetostriction and the compressibility share the same critical singularities. 

This latter finding has some interesting consequences. The diverging critical correction to the compressibility implies that the metamagnetic QCEP is intrinsically unstable with respect to a magnetoelastic coupling. Sufficiently close to the putative QCEP, the critical metamagnetic fluctuations will strongly renormalize the elastic constants and destabilize the crystal. We thus expect that the metamagnetic QCEP will be preempted by a structural instability whose properties will be the subject of a future publication.\cite{Garst09}

Our results are in particular relevant for itinerant ferromagnets with a tricritical point at zero field.\cite{Belitz05} In such a case there exists a metamagnetic QCEP at some finite $H_m$ whose critical thermodynamics might obey the predictions of spin-fluctuation theory. 
Such ferromagnets with a metamagnetic QCEP like for example UCoAl (Ref.~\onlinecite{Aoki11}) are thus promising candidates for a quantitative test of spin-fluctuation theory for critical metamagnetism.

\acknowledgments
We acknowledge useful discussions with L.~Fritz, T.~Lorenz, I.~Paul, A.~Rosch, and Y. Tokiwa, and an earlier collaboration with P.~Gegenwart and F.~Weickert that motivated this work. This work is supported by the DFG grants SFB 608 and FOR 960.

\appendix

\section{Sub-leading corrections}

\subsection{Quintic coupling}
\label{app:QuinticTerm}

The Ising symmetry of the potential (\ref{RenPotential}) emerges when the corrections due to quintic coupling $u_5$ can be neglected. At non-zero temperatures, the coupling $u_5$ will generate a cubic term which in turn renormalizes the linear coupling $h$. With these additional thermal corrections the expanded potential in the linear regime (\ref{EffPotExpanded})  obtains the modifications,
\begin{align}
\mathcal{V}(\phi) = \mathcal{V}(0) - H \phi + \frac{R}{2!} \phi^2 + \frac{U_3}{3!} \phi^3 + \frac{u}{4!} \phi^4 + \dots
\end{align}
where
\begin{align}
-H
&= -h + U_3 T^{(D-2)/z} \mathcal{A}'_d(R\, T^{-2/z})
\\
U_3
&= u_5  T^{(D-2)/z} \mathcal{A}'_d(R\, T^{-2/z}) 
\end{align}
with the same function $\mathcal{A}'_d$ as in the expression for the stiffness $R$, (\ref{EffMass}). 

In regime III, see Fig.~\ref{fig:regimes}, the resulting temperature correction to the tuning parameter, $h$, is of order $\mathcal{O}(u_5 T^{8/3})$ in $d=3$ and $\mathcal{O}(u_5 T^2 \log^2 T)$ in $d=2$. The correction is thus smaller ($d=3$) or of the same order (disregarding the logarithmic enhancement in $d=2$) as the intrinsic Fermi liquid temperature dependence that the parameter $h$ possesses anyway. These corrections to $h$ are thus not considered further. (The correction to thermodynamics due to an intrinsic $T^2$ depedence of $h$ are discussed in Section \ref{sec:IsingViolations}.) Moreover, it can be easily checked that the presence of the cubic term $U_3$ leads only to sub-leading corrections to the classical field configuration $\bar{\phi}$, see Eq.~(\ref{ClassicalField-Linear}).

In the non-linear regime, on the other hand, the quintic coupling can only be neglected at sufficiently small $h$. 
If the tuning $h$ reaches a threshold $|h| \gtrsim h_\Lambda$ with $h_\Lambda \sim u^{4}/u_5^{3}$, corrections to the scaling of Eq.~\eqref{ClassicalField-Nonlinear} become important and Ising symmetry is lost.
 
\subsection{Corrections to $\bar{\phi}$ in the non-linear regime}
\label{app:FieldCorrection}

In the derivation of the free energy in the non-linear regime (\ref{FreeEnergy-Nonlinear})
the zero temperature expression of the classical field configuration $\bar{\phi}$, Eq.~(\ref{ClassicalField-Nonlinear}), was used. Here we compute the corrections arising from the induced temperature correction to the classical field. Introducing the deviation $\delta \phi = \phi - \bar{\phi}$, the effective potential separates into two parts $\mathcal{V}(\phi) = \mathcal{V}(\bar{\phi}) + \delta\mathcal{V}(\delta\phi)$
with the correction
\begin{align}
\delta\mathcal{V}(\delta\phi) = \frac{u \bar{\phi}^2}{4} \delta\phi^2 
- f\, \delta \phi.
\end{align}
The temperature force is given by $f = F(T) - F(0)$ where
\begin{align}
F = - u \bar{\phi}\,  T^{(D-2)/z} \mathcal{A}'_d\left(\frac{u}{2} \bar{\phi}^2 T^{-2/z}\right) .
\end{align}
The temperature dependence of $F$ possesses similar limits as the one of $R$ in Eq.~(\ref{EffMassLimits}). By increasing the temperature one applies an effective force $f$ onto the field that modifies the position of the minimum. The resulting correction to the free energy is given by $\delta \mathcal{F}_{\rm cr} = - f^2/(u \bar{\phi}^2)$. This contribution is always sub-leading on the presented level of accuracy. 
 
\section{Asymptotic expansions}
\label{app:Details}

The asymptotic expansion of the function $\mathcal{A}_d$ is presented, see Eq.~(\ref{functionA}). Moreover, the numerical coefficients for the expansion of the potential's stiffness, $R$, Eq.~(\ref{EffMass}), and the free energy (\ref{FreeEnergyLimits}) are given. 

\begin{widetext}

\subsection{Dimension $d=3$}
\label{app:Dim3}

In spatial dimension $d=3$ the function $\mathcal{A}_d$ has the limiting behaviour
\begin{align}
\mathcal{A}_3(x) = 
\left\{
\begin{array}{ll}\displaystyle
 - \mathfrak{b}_1 \log\frac{\Lambda^3}{T} -\mathfrak{b}_0+ \mathfrak{b}_2 x - \mathfrak{b}_3 x^{\frac{3}{2}} + \dots
& {\rm for}\quad x\ll1
\\[0.5em]
\displaystyle
\bar{\mathfrak{b}}_1 x^{3} \log\frac{\Lambda^3}{T x^{3/2}} + \bar{\mathfrak{b}}_0  x^{3} - \bar{\mathfrak{b}}_2\, \log\frac{\Lambda^3}{T x^{3/2}} + \bar{\mathfrak{b}}_3\, x^{-3} \log x +\dots 
& {\rm for}\quad  \displaystyle \frac{\Lambda^2}{T^{2/3}} \gg x\gg 1.
\end{array}
\right.
\end{align}
There is a reminiscent dependence on the cutoff $\Lambda$. The coefficients
$\mathfrak{b}_0$ and $\bar{\mathfrak{b}}_0$ can be absorbed 
in the non-universal logarithmic dependence of the leading behavior, and both are omitted in the following.
The other coefficients read
\begin{align}
\mathfrak{b}_1 &= \frac{1}{36 \pi}, \quad
\mathfrak{b}_2 = \frac{1}{6 \sqrt{3} \pi^2} \Gamma(4/3) \zeta(4/3), \quad
\mathfrak{b}_3 = \frac{1}{12\pi}, \quad
\bar{\mathfrak{b}}_1 = \frac{1}{72 \pi^3},\quad
\bar{\mathfrak{b}}_2 = \frac{1}{36 \pi},\quad 
\bar{\mathfrak{b}}_3 = \frac{\pi}{60}.
\end{align}
The leading temperature correction to the stiffness is given by
\begin{align}
R = 
\left\{
\begin{array}{ll}
r + \mathfrak{b}_2\, u\, T^{4/3} &  \quad {\rm for}\quad  R \ll T^{2/3}
\\
r + \frac{3}{2}\bar{\mathfrak{b}}_2\, u\,\frac{T^2}{r} & \quad  {\rm for}\quad R \gg T^{2/3}.
\end{array}
\right.
\end{align}
The limiting form of the free energy in the four different regimes of Fig.~\ref{fig:regimes}
reads
\begin{align}
\mathcal{F}_{\rm cr} = 
& \left\{
\begin{array}{cc}
\displaystyle 
- \bar{\mathfrak{b}}_2\, T^2
\log \frac{\sqrt{2}\Lambda^3}{3 u^{1/2} |h| }
+ \bar{\mathfrak{b}}_3 \frac{2}{9} \frac{T^4}{u h^2} 
\log  \frac{|h|^{2/3} u^{1/3}}{T^{2/3}}
- \frac{(3 |h|)^{4/3}}{2^{5/3} u^{1/3}}
& \quad{\rm I}\\[1em]
\displaystyle
-\mathfrak{b}_1 T^2 \log\frac{\Lambda^3}{T} 
+ \mathfrak{b}_2 \frac{6^{2/3}}{2} T^{4/3} (u h^2)^{1/3} 
- \frac{(3 |h|)^{4/3}}{2^{5/3} u^{1/3}}
& \quad{\rm II}\\[1em]
\displaystyle
-\mathfrak{b}_1 T^{2} \log \frac{\Lambda^3}{T} 
- \frac{h^2}{2 R}
& \quad{\rm III}\\[1em]
\displaystyle
- \bar{\mathfrak{b}}_2 T^2 \log\frac{\Lambda^3}{r^{3/2}} 
+ \bar{\mathfrak{b}}_3 \frac{T^4}{r^3} \log \frac{r}{T^{2/3}}
- \frac{h^2}{2 R}
& \quad{\rm IV}
\end{array}
\right.
\end{align}

\subsection{Dimension $d=2$}
\label{app:Dim2}   

In spatial dimension $d=2$ the asymptotic form of the function $\mathcal{A}_d$ is
\begin{align}
\mathcal{A}_2(x) = 
\left\{
\begin{array}{ll}\displaystyle
- \mathfrak{c}_1 + \mathfrak{c}_3 x \log\frac{1}{x}
+ \dots
& {\rm for}\quad x\ll1
\\[0.5em]\displaystyle
\bar{\mathfrak{c}}_1\,
x^{\frac{5}{2}} - \bar{\mathfrak{c}}_2\, x^{-\frac{1}{2}} +
\bar{\mathfrak{c}}_3 x^{-2} + \dots 
&  {\rm for}\quad x\gg 1
\end{array}
\right.
\end{align}
The function is universal in the limit $\Lambda/T^{1/3} \to \infty$.
The coefficients are
\begin{align}
\mathfrak{c}_1 &= \frac{1}{4 \pi} \Gamma(5/3) \zeta(5/3),\quad
\mathfrak{c}_3 = \frac{1}{8\pi},\quad
\bar{\mathfrak{c}}_1 = \frac{1}{30 \pi},\quad
\bar{\mathfrak{c}}_2 = \frac{\pi}{24} ,\quad
\bar{\mathfrak{c}}_3 = \frac{1}{4 \pi}\zeta(3) .
\end{align}
The term with coefficient $\mathfrak{c}_3$ derives from the zero Matsubara mode whose contribution is logarithmically enhanced in $d=2$. This logarithmic dependence  is reflected in the temperature dependence of the  stiffness
\begin{align}
R = 
\left\{
\begin{array}{ll}
r + \mathfrak{c}_3\, u\, T \log \frac{T^{2/3}}{R} &  \quad {\rm for}\quad  R \ll T^{2/3}
\\
r + \frac{1}{2}\bar{\mathfrak{c}}_2\, u\,\frac{T^2}{r^{3/2}} & \quad  {\rm for}\quad R \gg T^{2/3}.
\end{array}
\right.
\end{align}
For $T \gg r/u$ in the quantum critical regime, the temperature determines the stiffness so that asymptotically $R = \mathfrak{c}_3\, u\, T \log (1/(u T^{1/3}))$. The free energy simplifies in the four regimes of Fig.~\ref{fig:regimes} to 
\begin{align} \label{FreeEnergy2d}
\mathcal{F}_{\rm cr} = 
& \left\{
\begin{array}{cc}
\displaystyle
- \bar{\mathfrak{c}}_2\,\frac{2^{1/2}}{6^{1/3}} \frac{T^2}{(u h^2)^{1/6}} 
+ \bar{\mathfrak{c}}_3 \frac{4}{6^{4/3}} \frac{T^3}{(u h^2)^{2/3}}  
- \frac{(3 |h|)^{4/3}}{2^{5/3} u^{1/3}}
& \quad{\rm I}\\[1em]
\displaystyle
-\mathfrak{c}_1 T^{5/3} 
+ \mathfrak{c}_3 \frac{6^{2/3}}{2} T (u h^2)^{1/3} \log \frac{T^{2/3}}{(u h^2)^{1/3}}
- \frac{(3 |h|)^{4/3}}{2^{5/3} u^{1/3}}
& \quad{\rm II}\\[1em]
\displaystyle
-\mathfrak{c}_1 T^{5/3} 
- \frac{h^2}{2 R}
& \quad{\rm III}\\[1em]
\displaystyle
- \bar{\mathfrak{c}}_2\,\frac{T^2}{r^{1/2}} 
+ \bar{\mathfrak{c}}_3 \frac{T^3}{r^2}  
 - \frac{h^2}{2 R}
& \quad{\rm IV}
\end{array}
\right.
\end{align}
\end{widetext}
The terms of the free energy with coefficients $\bar{\mathfrak{b}}_3$ and $\bar{\mathfrak{c}}_3$ that lead to non-analytic Fermi liquid corrections are omitted in the body of the paper.

\end{document}